\input{aipcheck}

\documentclass[
    ,final            
  ]
  {aipproc}

\layoutstyle{6x9}

\begin{document}

\title{New Calculations of Stellar Wind Torques}



\author{Sean P. Matt}{
  address={Department of Astronomy, University of Virginia, P.O. Box 400325, Charlottesville, VA 22904},altaddress={Current address: NASA Ames Research Center, MS 245-6, Moffett Field, CA 94035-1000}
}

\author{Ralph E. Pudritz}{
  address={Physics \& Astronomy Department, McMaster University, Hamilton, ON L8S 4M1, Canada}
}

\begin{abstract}

Using numerical simulations of magnetized stellar winds, we carry out
a parameter study to find the dependence of the stellar wind torque on
observable parameters.  We find that the power-law dependencies of the
torque on parameters is significantly different than what has been
used in all spin evolution models to date.

\end{abstract}

\maketitle


\newcommand{\aap}   {A\&A}
\newcommand{\aj}    {AJ}
\newcommand{\apj}   {ApJ}
\newcommand{\apjl}  {ApJ}
\newcommand{\araa}  {ARA\&A}
\newcommand{\mnras} {MNRAS}
\newcommand{\nat}   {Nature}
\newcommand{\pasj}  {PASJ}
\newcommand{\pasp}  {PASP}


\section{Introduction}

Studying rotation gives us a better understanding of how stars form,
how they function, how they evolve, and how they interact with their
environment.  To understand the rotation rates and rotational
evolution of cool stars, we must understand how they lose angular
momentum.  We know that stellar winds are of primary importance for
this, but the existing analytic theory for calculating the stellar
wind torque has not substantially changed in 20 years.  Since that
time, numerical simulations have revealed that a number of the
underlying assumptions of this theory are not generally valid and that
a new formula for the torque is needed.

Numerical simulations have the advantage that they typically require
fewer assumptions (and often fewer tunable parameters) than analytic
theory.  However, the disadvantage is that a single simulation does
not give much insight into how the physics depends upon the
parameters.  To improve this situation, \citet{mattpudritz08II}
carried out a small simulation parameter study, using ideal
magnetohydrodynamic simulations to compute axisymmetric, steady-state
solutions of Solar-like winds from magnetized stars.  That work
focused on accretion-powered winds from pre-main-sequence stars, and
the reader will find all details in that paper.  Here, we summarize
the results and rescale the physical parameters and results of those
simulations to be in a regime more representative of Solar-like main
sequence stars.

\section{Wind Simulation Method and Results}

The parameters and values adopted for the fiducial wind are listed in
table \ref{tab_parms}.  The parameters are the stellar mass ($M_*$),
radius ($R_*$), dipole magnetic field strength at the equator ($B_*$),
wind mass outflow rate ($\dot M_{\rm w}$), wind thermal sound speed
divided by the surface escape speed ($c_{\rm s} / v_{\rm esc}$), the
adiabatic index ($\gamma$, where $P \propto \rho^\gamma$), and the
spin rate expressed as a fraction of breakup speed, $f \equiv \Omega_*
R_*^{3/2} (G M_*)^{-1/2}$, where $\Omega_*$ is the angular spin rate
of the star and $G$ is Newton's gravitational constant.


\begin{table}
\begin{tabular}{ll}
\hline

 \tablehead{1}{l}{b}{Parameter}
& \tablehead{1}{l}{b}{Value} \\

\hline

$M_*$                     & 1.0 $M_\odot$ \\
$R_*$                     & 1.0 $R_\odot$ \\
$B_*$ (dipole)            & 10 Gauss     \\
$\dot M_{\rm w}$           & $5.9 \times 10^{-13} M_\odot$/yr \\
$f$                       & 0.1           \\
$c_{\rm s} / v_{\rm esc}$   & 0.222          \\
$\gamma$                  & 1.05           \\

\hline

\end{tabular}

\caption{Fiducial Stellar Wind Parameters \label{tab_parms}}

\end{table}

In the simulations, the wind solution is completely determined by the
boundary conditions applied at the surface of the star.  At that
boundary, the magnetic field is anchored into the star, which is
assumed to be rotating as a solid body.  The field geometry is a
dipole at the stellar surface (except for two cases with a quadrupolar
field).  The surface temperature and density is uniform and held fixed
throughout the simulation.  The wind velocity is determined
dynamically by the simulations.

We ran the fiducial case and 16 other cases, listed in
table \ref{tab_results}.  In each case, we changed only one parameter
value relative to the fiducial case.  The changed parameter and its
value is listed in the first column of table \ref{tab_results}.

\begin{table}[h]
\begin{tabular}{lccc}
\hline

 \tablehead{1}{l}{b}{Case} 
& \tablehead{1}{c}{b}{$\dot M_{\rm w}$\\($10^{-13} {M_\odot \over {\rm yr}}$)}
& \tablehead{1}{c}{b}{$\tau_{\rm w}$\\($10^{32}$ ${\rm erg}$)} 
& \tablehead{1}{c}{b}{$r_{\rm A}$\\$(R_*)$} \\

\hline

fiducial                     &  5.91  &  5.53   &  6.97  \\
$f$ = 0.004                  &  5.81  &  0.304  &  8.33  \\
$f$ = 0.2                    &  5.84  &  8.81   &  6.26  \\
$f$ = 0.05                   &  5.88  &  3.31   &  7.65  \\
$B_*$ = 20 G                 &  5.81  &  10.2   &  9.55  \\
$B_*$ = 100 G                &  6.00  &  43.1   &  19.3  \\
50 G quadrupole                   &  5.84  &  4.28   &  6.17  \\
100 G quadrupole                  &  6.03  &  2.11   &  7.53  \\
low $\dot M_{\rm w}$          &  0.584 &  1.56   &  11.8  \\
very low $\dot M_{\rm w}$     &  0.118 &  0.638  &  16.7  \\
$R_*$ = 0.75 $R_\odot$        &  5.81  &  3.44   &  5.96  \\
$R_*$ = 1.5 $R_\odot$         &  5.91  &  10.7   &  8.75  \\
$M_*$ = 0.5 $M_\odot$         &  5.97  &  4.59   &  7.52  \\
$M_*$ = 2 $M_\odot$           &  5.88  &  6.59   &  6.42  \\
$c_{\rm s}/v_{\rm esc}$ = 0.245 &  5.84  & 4.97  &  6.64    \\
$c_{\rm s}/v_{\rm esc}$ = 0.192 &  5.91  & 5.97  &  7.23   \\
$\gamma$ = 1.10              &  5.84  &  6.84   &  7.79  \\

\hline

\end{tabular}

\caption{Simulated Wind Results \label{tab_results}}

\end{table}

The simulations result in a Parker-like, thermally driven wind that is
modified, self-consistently, by the stellar rotation, presence of the
magnetic field, and polytropic equation of state.  Thus, the total
mass loss rate in the flow, $\dot M_{\rm w}$ listed in
table \ref{tab_results}, is not a true parameter but is actually a
result of the simulation.  In order to treat $\dot M_{\rm w}$ as a
pseudo-parameter, our method is to iteratively change the value of the
density at the stellar surface until the desired value of $\dot M_{\rm
w}$ is obtained, within a tolerance of a few percent.

After each simulation reaches a steady-state solution, we calculate
the total angular momentum loss rate, $\tau_{\rm w}$, by integrating
the angular momentum flux over a surface enclosing the star.  Then we
use the formula
\begin{equation}
\label{eq_tau}
\\
\tau_{\rm w} = \dot M_{\rm w} \Omega_* r_{\rm A}^2,
\\
\end{equation}
from analytic theory, to calculate the value of $r_{\rm A}$, the
``magnetic lever arm'' length.  Since the winds are multi-dimensional,
$r_{\rm A}$ represents the mass-loss-weighted average of the Alfv\'en
radius in the flow.  Table \ref{tab_results} lists the results for each
case.

\section{Toward a Predictive Torque Theory}

\begin{figure}
  \includegraphics[height=.35\textheight]{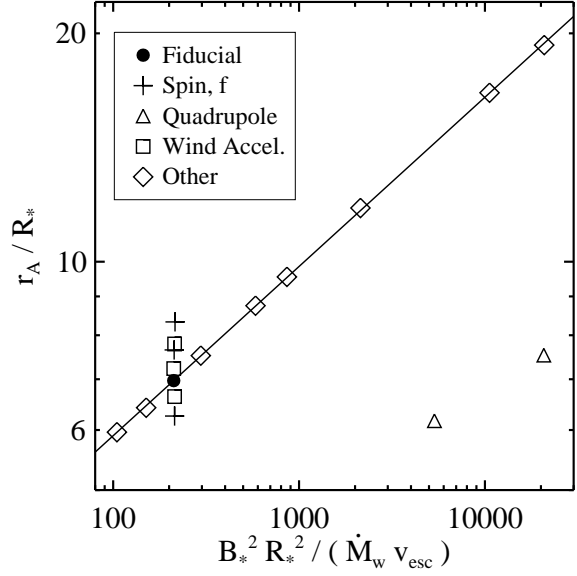} 

\caption{Effective magnetic lever arm length, in units of stellar radii, 
versus the dimensionless combination of parameters $B_*^2 R_*^2 (\dot
M_{\rm w} v_{\rm esc})^{-1}$.  The cases with variations in $B_*$,
$R_*$, $\dot M_{\rm w}$, or $M_*$ are shown with {\it diamonds}, while
other symbols represent the remaining cases.  The line shows the best
fit to the {\it diamonds} and the fiducial case, given by equation
(\ref{eq_ra}) with $K \approx 2.11$ and $m \approx 0.223$.  Figure
from \citep{mattpudritz08II}.}

\label{fig_raplot}
\end{figure}

Analytic theory suggests that the dimensionless combination of
parameters $B_*^2 R_*^2 (\dot M_{\rm w} v_{\rm esc})^{-1}$ is of
fundamental importance for determining the wind physics.  We plot
$r_{\rm A} / R_*$ as a function of this parameter combination in
figure \ref{fig_raplot}.  Using a simple power law formulation,
\begin{equation}
\label{eq_ra}
\\
{r_{\rm A} \over R_*} = K \left({B_*^2 R_*^2 \over \dot M_{\rm w} v_{\rm esc}}\right)^m,
\\
\end{equation}
we find that $K \approx 2.11$ and $m \approx 0.223$ provides an
excellent fit to the fiducial case and the simulations with variations
in $B_*$, $R_*$, $\dot M_{\rm w}$, or $M_*$.

The combination of equations (\ref{eq_tau}) and (\ref{eq_ra}) gives
\begin{equation}
\label{eq_total}
\\
\tau_{\rm w} = K^2 {R_*^{5m+2} \over (2 G M_*)^{m}} 
              \Omega_* B_*^{4m} \dot M_{\rm w}^{1-2m}.
\\
\end{equation}
This is essentially the same formula as derived by \citet{kawaler88}.
However, without the aid of numerical simulations, Kawaler had to
parameterize the magnetic field structure in the wind and make
assumptions about how the wind velocity depends on parameters.  Thus,
he preferred a value of $m = 0.5$, which is significantly different
than the value obtained by our simulations.  The value obtained by
simulations should be viewed as more accurate, since the simulations
self-consistently calculate the field structure and wind velocity in
multiple dimensions and thus have fewer parameters and assumptions.

Finally, it is important to point out that neither the work presented
here nor any previous analytic theory has properly determined the
dependence of the torque on the spin rate or wind driving parameters.
This is evident in figure \ref{fig_raplot}, since the fit of equation
(\ref{eq_ra}) is less precise for cases with different spin rates
($f$) or wind acceleration parameters ($c_{\rm s}/v_{\rm esc}$ or
$\gamma$).  Also, it is clear from the outlying {\it triangles} in the
figure that the torque is very sensitive to field geometry.  Due to
the limited number of simulations in this study with changes in the
spin rate, wind driving, and field geometry, it is still not clear
what is the appropriate functional form for the dependence of $r_{\rm
A}$ on these parameters.  Future work will include a larger parameter
study, with the goal of determining the precise dependence on all
relevant parameters.

This work has implications for understanding the observations of
stellar rotation at all evolutionary phases, as well as the
emperically established method of gyrochronology \citep{barnes03}.  As
an illustrative example of how the new power law index $m$ changes our
understanding, consider that the empirical \citet{skumanich72}
relationship for the spin-down of main-sequence stars suggests
$\tau_{\rm w} \propto \Omega_*^3$.  Under the assumption that $m =
0.5$ in equation (\ref{eq_total}), a particular dynamo relationship,
where $B_* \propto \Omega_*$, elegantly explains the Skumanich
relationship.  However, we showed that $m$ is instead much closer to
0.2.  This suggests, e.g., either that a different dynamo relationship
is appropriate or that the mass loss rate depends in a particular way
on the stellar rotation rate.

\begin{theacknowledgments}

The workshop organizers deserve credit for an excellent conference.
The work of SPM was supported by the University of Virginia through a
Levinson/VITA Fellowship partially funded by the Frank Levinson Family
Foundation through the Peninsuly Community Foundation.  REP is
supported by a grant from NSERC.

\end{theacknowledgments}






\begin{thebibliography}{4}
\expandafter\ifx\csname natexlab\endcsname\relax\def\natexlab#1{#1}\fi
\providecommand{\enquote}[1]{``#1''}
\expandafter\ifx\csname url\endcsname\relax
  \def\url#1{\texttt{#1}}\fi
\expandafter\ifx\csname urlprefix\endcsname\relax\def\urlprefix{URL }\fi
\providecommand{\eprint}[2][]{\url{#2}}

\bibitem[{Matt} and {Pudritz}(2008)]{mattpudritz08II}
S.~{Matt}, and R.~E. {Pudritz}, \emph{\apj} \textbf{678}, 1109--1118 (2008),
  \eprint{arXiv:0801.0436}.

\bibitem[{Kawaler}(1988)]{kawaler88}
S.~D. {Kawaler}, \emph{\apj} \textbf{333}, 236--247 (1988).

\bibitem[{Barnes}(2003)]{barnes03}
S.~A. {Barnes}, \emph{\apj} \textbf{586}, 464--479 (2003).

\bibitem[{Skumanich}(1972)]{skumanich72}
A.~{Skumanich}, \emph{\apj} \textbf{171}, 565 (1972).

\end{thebibliography}

\end{document}